\documentstyle[psfig,12pt]{article}
\begin{document}
{\baselineskip12pt
\vspace*{0.4cm}
\begin{center}
\begin{Large}
ANALYSIS OF THE THERMAL CROSS SECTION OF THE CAPTURE REACTION
$^{13}$C(n,$\gamma$)$^{14}$C 
\end{Large}
\end{center}
\vspace{3mm}
\begin{center}
H. Herndl and H. Oberhummer\\
Institut f\"ur Kernphysik, TU Wien,\\
Wiedner Hauptstr. 8--10, A--1040 Vienna, Austria\\
\vspace{4mm}
\end{center}

{\it Abstract:\/} We investigate the thermal cross section of the
reaction
$^{13}$C(n,$\gamma$)$^{14}$C which takes place in the helium
burning zones of red giant star as well as in the
nucleosynthesis of Inhomogeneous Big Bang models.
We find that we can reproduce the experimentally
known thermal capture cross section only if we take into
account a strong hindrance of the E1 transition in the
nuclear interior. This effect can be explained by the strong
coupling to the giant dipole resonance.
\baselineskip18pt
\vspace{7mm}

\section{Introduction}
The neutron capture reaction $^{13}$C(n,$\gamma$)$^{14}$C plays an
important role in nuclear astrophysics.
In stellar helium burning it can act as neutron poison for the slow 
neutron capture process (s--process). 
As one of the main products of the CNO--cycle $^{13}$C is also abundant in
helium burning zones. In low mass asymptotic giant branch (AGB) stars
the reaction $^{13}$C($\alpha$,n)$^{16}$O is considered to be the
main source of neutron production. In this case the capture
$^{13}$C(n,$\gamma$)$^{14}$C can reduce not only the neutron but
also the $^{13}$C abundance \cite{ibe76}.

Furthermore the reaction is also important in the nucleosynthesis
of Inhomogeneous Big Bang models. In neutron--rich zones
intermediate--mass nuclei might be produced via a reaction
sequence passing through the neutron capture reactions
$^{12}$C(n,$\gamma$)$^{13}$C(n,$\gamma$)$^{14}$C
\cite{kaj90}.

In Section 2 we will present the methods used to  
determine the reaction cross section. Then we
will discuss the experimental and theoretical input parameters for
our calculations in Section 3. 
In Section 4 we will consider the characteristics of the s--wave capture
which will lead us to the use of a semidirect model. Finally we will  
summarize and discuss our results in Section 5. 

\section{Calculation of the Cross Section}

The thermal cross section is, if there is no resonant contribution,
given by direct s--wave capture. We 
calculate the direct capture (DC) cross section in a potential model
described in\,\cite{kim87,obe91,moh93}. The total nonresonant cross
section $\sigma_{\rm nr}$ is determined by the direct capture transitions
$\sigma^{\rm DC}_i$ to all bound states with the single particle
spectroscopic factors $C^2 S_i$ in the final nucleus:
\begin{equation}
\label{NR}
\sigma^{\rm DC}_{\rm tot} = \sum_{i} \: (C^{2} S)_i\sigma^{\rm DC}_i \quad .
\end{equation}
The DC cross sections $\sigma^{\rm DC}_i$ are
determined by the overlap of the scattering wave function
in the entrance channel, the bound--state wave function
in the exit channel and the multipole transition--operator.
 
The most important ingredients in the potential model are the wave functions 
for the scattering and bound states in the entrance and exit channels.
For the calculation of these wave functions we use real folding potentials
which are given by\,\cite{obe91,kob84}
\begin{equation}
\label{FO}
V(R) = 
  \lambda\,V_{\rm F}(R) 
  = 
  \lambda\,\int\int \rho_a({\bf r}_1)\rho_A({\bf r}_2)\,
  v_{\rm eff}\,(E,\rho_a,\rho_A,s)\,{\rm d}{\bf r}_1{\rm d}{\bf r}_2 \quad ,
\end{equation}
with $\lambda$ being a potential strength parameter close
to unity, and $s = |{\bf R} + {\bf r}_2 - {\bf r}_1|$,
where $R$ is the separation of the centers of mass of the
projectile and the target nucleus.
The density can been derived from measured
charge distributions\,\cite{vri87}
and the effective nucleon--nucleon
interaction $v_{\rm eff}$
has been taken in the DDM3Y parametrization\,\cite{kob84}.
The imaginary part of the potential 
is very small because of the small flux into other reaction channels
and can be neglected in most cases involving neutron capture
by neutron--rich target nuclei.

\section{Determination of the Nuclear Input Parameter}

The input parameters for direct s--wave capture transitions of the
reaction $^{13}$C(n,$\gamma$)$^{14}$C are listed in
Table~\ref{tab-dcc}.
The spectroscopic factors are important input parameters. They can be obtained 
experimentally from single--particle transfer reaction studies. 
In our case we have used the  
spectroscopic factors obtained from the reaction
$^{13}$C(d,p)$^{14}$C \cite{ajz86a} for the s--wave transitions to
the $0^+$ state at 6.59 MeV and to the $2^+$ state at 7.01 MeV in $^{14}$C.
The spectroscopic factor for the s--wave transition to the ground
state of $^{14}$C is taken from a shell--model calculation \cite{coh67}.

\begin{table}[htb]
\caption{Considered s--wave transitions for the direct contribution to the 
reaction $^{13}$C(n,$\gamma$)$^{14}$C.     
The spectroscopic factor for the ground state transition is taken
from the shell model calculation of Ref.~[10], while the
spectroscopic factors for the other transitions are from a
$^{13}$C(d,p)$^{14}$C experiment [9].}
\vspace{4mm}
\begin{center} 
\begin{tabular}{|rccrr|}
\hline
 $J^{\pi}$ & $E_x$ (MeV) & Q--value (MeV) & transition & $C^2 S$ \\
\hline
 $0^+$ & 0.000 & 8.176 & s$\to$1p$_{1/2}$ &
1.734 \\
 $0^+$ & 6.589 & 1.587 & s$\to$2p$_{1/2}$ & 0.14 \\
 $2^+$ & 7.012 & 1.164 & s$\to$2p$_{3/2}$ & 0.065 \\
\hline
\end{tabular}
\end{center}
\label{tab-dcc}
\end{table}

%\begin{figure}
%\centerline{\psfig{figure=Fig1.eps,width=13cm}}
%\caption{\label{f1} Level density (in levels per MeV) at the
%respective neutron separation
%energy \protect\cite{Rau95b,Rau96}.}
%\end{figure}

The potential strength parameter 
$\lambda$ for the bound state is adjusted to the
known binding energies. In the scattering state we determine
$\lambda$ from the scattering lengths. For neutron scattering on $^{13}$C
the free scattering lengths are given by 
$a_{\rm c}=5.76$fm and $a_{\rm i}=-0.48$fm, where
the subscript c and i denote the coherent and incoherent
scattering length, respectively. Since the incoherent scattering is not
negligible in this case we determine two different potentials for the channel
spins $J_{\rm a}=0$ and $J_{\rm a}=1$, which are $\lambda_0=1.1396$ and
$\lambda_1=1.2465$, respectively.

\begin{figure}
\centerline{\psfig{figure=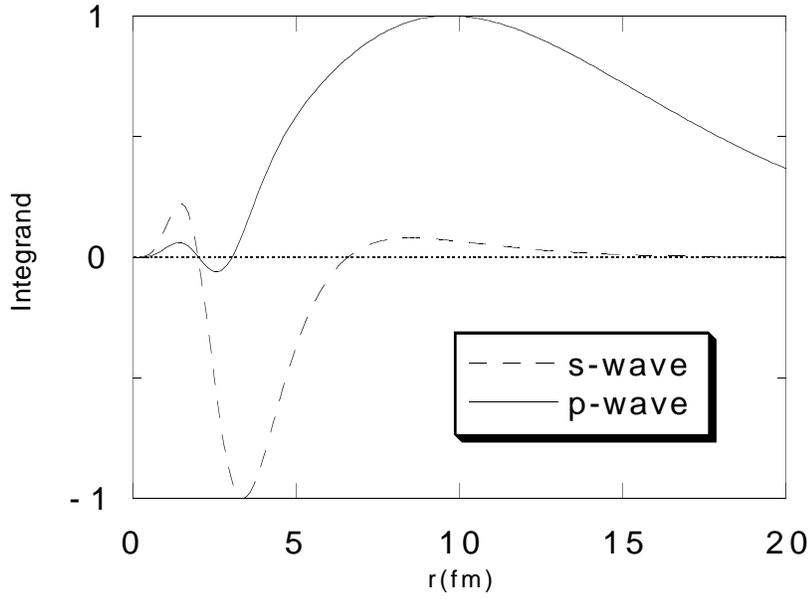,width=13cm}}
\caption[fig1]{The radial integrand for the direct capture of thermal 
s--wave capture and p--wave capture at the thermonuclear energy
$E=10$ keV.}
\label{fig-int}
\end{figure}

\begin{figure}
\centerline{\psfig{figure=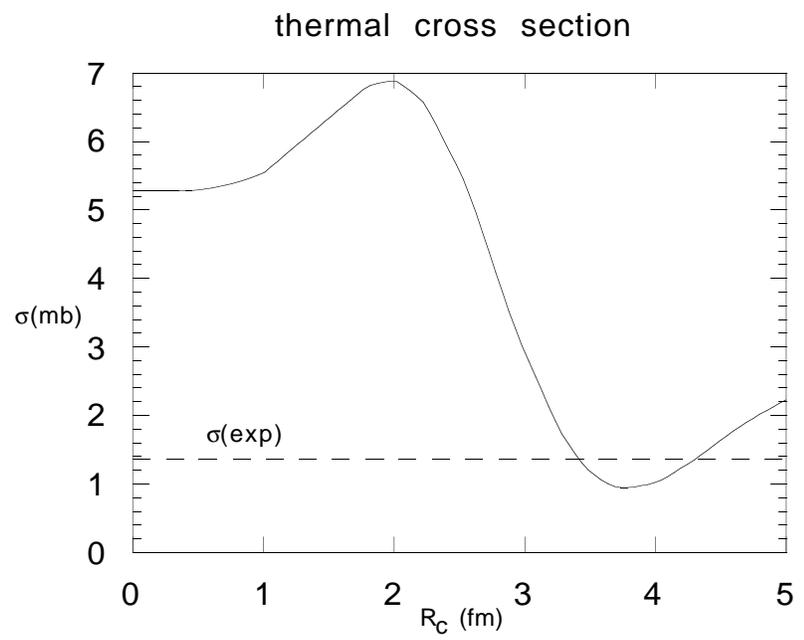,width=13cm}}
\caption[fig2]{The dependence of the thermal s--wave cross section on the
cutoff radius $R_c$ in the cutoff model. We compare the cross
section with the known thermal cross section of 1.37 mb.}
\label{fig-sds}
\end{figure}

\begin{figure}
\centerline{\psfig{figure=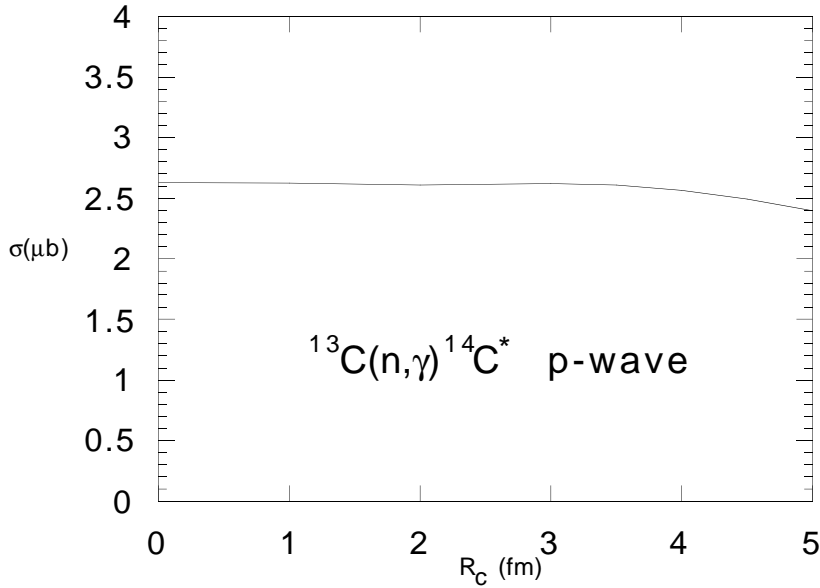,width=13cm}}
\caption[fig3]{Dependence of the thermonuclear p--wave capture cross
section on the cutoff radius $R_c$ in the cutoff model.}
\label{fig-sdp}
\end{figure}
\section{The Thermal Capture Cross Section}

The thermal s--wave capture cross section is clearly dominated
by the transition to the ground state of $^{14}$C. The other two
transitions are negligible due to the much lower spectroscopic
factors and the higher excitation energies.
Using the data of Table~\ref{tab-dcc} we obtain a thermal 
capture cross
section of 5.28 mb, clearly higher than the measured value of 1.37 mb.
In order to understand this discrepancy we look at the direct--capture
radial integrand of
this transition. From Fig.~\ref{fig-int} we see that major contributions come
from the nuclear interior. The maximum of the integrand is at 
around 3.4 fm which
is close to the surface of the nucleus. We compare this with the 
integrand of the p--wave
transition to the first excited state at an incident energy of 10 keV. In this
case the most important contributions come from the nuclear exterior. 
The maximum
of the integrand lies at approximately 10 fm.

In the nuclear interior both the scattering and the bound state wave function 
are modified due to the coupling to the giant dipole resonance. 
This coupling causes a hindrance of the E1 capture cross section at
low energies. These effects were discussed by several authors,
usually at higher energies \cite{zim70,uch85}. If the neutron energies
is close to the energy of the giant diploe resonance the coupling is
attractive causing an increase of the cross section.
In order to estimate the influence of this effect on our reaction
we use the simple
cutoff model from Ref.~\cite{uch85}. In this model an effective
charge
\begin{equation}
e_{\rm eff}(r) = e_{\rm E1} \theta (r-R_c)
\end{equation}
where $e_{\rm E1}$ is the equal to $-Ze/A$ for neutron capture
and $R_c$ is a cutoff radius. Inside this cutoff radius the
effective charge vanishes.

In Fig.~\ref{fig-sds} the dependence of the s--wave capture
cross section at thermal energies on the cutoff radius is
shown. While the cross section is clearly too high for a small
cutoff radius, it is very close to the experimental thermal cross
section of 1.37 mb if the cutoff radius is around 4 fm.
This is very near to the nuclear radius. On the other hand
the thermonuclear p--wave capture does not show a strong
dependence on the cutoff radius (see Fig.~\ref{fig-sdp}).

\section{Summary and Discussion}

We have shown that the thermal cross section of the reaction
$^{13}$C(n,$\gamma$)$^{14}$C can be well described in a potential
model if one takes into account the coupling of the wave functions
to the giant dipole resonance in the nuclear interior. This effect
causes a strong hindrance of the E1 capture cross sections if the
main contributions to the direct capture integrand come from the
nuclear interior. This is especially the case for incoming
s--waves.

%\acknowledgments
%This work was supported by
%Fonds zur F\"orderung der wissenschaftlichen Forschung 
%(FWF project S7307--AST).

%

%
\end{document}